\documentclass{article}

\usepackage{fullpage}
\usepackage{latexsym}
\usepackage{amsmath}
\usepackage{amssymb}
\usepackage{amsfonts}

\def\01{\{0,1\}}

\newcommand{\eps}{\varepsilon}
\newcommand{\Exp}{\mbox{\rm E}}

\newcommand{\ket}[1]{|#1\rangle}

\newtheorem{theorem}{Theorem}
\newtheorem{lemma}[theorem]{Lemma}

\newtheorem{corollary}[theorem]{Corollary}

\newtheorem{claim}[theorem]{Claim}

\newenvironment{proof}
{\noindent {\bf Proof. }}
{{\hfill $\Box$}\\
 \smallskip}

\begin{document}

\title{Exponential Separation of Quantum and Classical One-Way Communication Complexity for a Boolean Function}
\author{Dmitry Gavinsky\\University of Calgary
 \and Julia Kempe\thanks{Supported in part by ACI S\'ecurit\'e Informatique
SI/03 511 and ANR AlgoQP grants of the French Research Ministry,
and also partially supported by the European Commission under the
Integrated Project Qubit Applications (QAP) funded by the IST
directorate as Contract Number 015848.}\\CNRS \&\ LRI\\ Univ.~de
Paris-Sud, Orsay
 \and Ronald de Wolf\thanks{Supported by a Veni grant
from the Netherlands Organization for Scientific Research (NWO)
and also partially supported by the European Commission under the
Integrated Projects Qubit Applications (QAP) funded by the IST
directorate as Contract Number 015848.}\\CWI, Amsterdam }
\date{}
\maketitle

\begin{abstract}
We give an exponential separation between one-way quantum and classical
communication complexity for a Boolean function.  Earlier such a separation was
known only for a relation.  A very similar result was obtained earlier but
independently by Kerenidis and Raz~\cite{kerenidisraz:forafunction}.  Our
version of the result gives an example in the bounded storage model of cryptography,
where the key is secure if the adversary has a certain amount of classical storage,
but is completely insecure if he has a similar amount of \emph{quantum} storage.
\end{abstract}

\section{Introduction}

From a computer science perspective, the main theoretical goal of the field of
quantum computing is to exhibit problems where quantum computers are much
faster (or otherwise better) than classical computers. Preferably exponentially
better. The most famous example, Shor's efficient quantum factoring algorithm,
constitutes a separation only if one is willing to believe that efficient
factoring is impossible on a classical computer---proving this would of course
imply that P$\neq$NP. One of the few areas where we can establish
\emph{unconditional} exponential separations is the area of communication
complexity.  Here two parties, Alice with input $x$ and Bob with input $y$,
collaborate to solve some computational problem that depends on both $x$ and
$y$. They want to do this with minimal communication.

Examples of communication problems where quantum communication gives
exponential savings over classical communication were for instance given by
Buhrman, Cleve, and Wigderson~\cite{BuhrmanCleveWigderson98} for zero-error protocols, Raz~\cite{raz:qcc} for bounded-error protocols, Buhrman, Cleve, Watrous, and de Wolf~\cite{bcww:fp} for simultaneous message passing protocols, and
Bar-Yossef, Jayram, and Kerenidis~\cite{bjk:q1way} for one-way protocols. The last result establishes
an exponential separation for one-way communication: it describes a problem (the Hidden Matching Problem)
which can be solved by a quantum message of $\log n$ qubits, but which cannot
be solved with good success probability with much fewer than $\sqrt{n}$
classical bits of communication. However, their problem is a \emph{relational
problem}, where for each $x$ and $y$ many possible outputs are considered
correct. Establishing a separation for a Boolean \emph{function} was left as an
open problem.

In this paper, we give an exponential ($(\log n)^{3/2}$ vs $\sqrt{n}(\log
n)^{1/4}$) quantum-classical separation for one-way communication for a
Boolean function. The problem is a variant of a functional problem that was
already conjectured to give such a separation by Bar-Yossef et al. As usual in
such results, the efficient quantum protocol is quite easy, while the lower
bound for classical one-way protocols is harder to show. A very similar
separation was obtained earlier but independently by Kerenidis and Raz~\cite{kerenidisraz:forafunction} using different techniques based on Fourier analysis. The outline of their proof was obtained in the summer of
2004, but was written up in detail when they learned in June 2006 about
our independent work (which was not quite finished at the time). We thank
Kerenidis and Raz for generously delaying the publication of their proof while
we were finishing ours.

Let us briefly point out some differences between our two proofs. The
Kerenidis-Raz result is a slightly stronger separation than ours, since the
quantum upper bound for their problem is $\log n$ while ours is $(\log
n)^{3/2}$. Their proof is heavily based on the analysis of Fourier
coefficients. It is self-contained except for an application of the
Bonami-Beckner inequality. Our proof is quite different. Intuitively, it shows
that if Alice's message was too short, then Bob has hardly any information
about a certain string $z$ that can be computed from $x$ given also Bob's
input. It is based on a result of Talagrand \cite{talagrand:increasingsets}
and a large deviation inequality for martingales due to McDiarmid~\cite{McDiarmid:concentration}.
A small advantage of our result is that our
quantum protocol has zero-error, while the Kerenidis-Raz variant of the problem gives a
separation between bounded-error quantum and classical, but not between
zero-error quantum and bounded-error classical.

Another advantage of our proof is that it shows something interesting about the
bounded storage model of cryptography.  In this model, introduced by Maurer~\cite{Maurer92b}, an adversary has full but temporary access to some string $x$
but can only store a limited amount of information about $x$.  A and B have a
secret key $K$ available, which they use to derive a common string $z$ from $x$
(they only need to store small portions of $x$ at the time). This string $z$ is
supposed to be almost completely uniformly distributed from the point of view of
the adversary, and hence can be used by A and B as a key in a one-time pad communication scheme. The power of this
model is that one can show in many cases that the adversary knows very little
about the string $z$ that Alice and Bob derived from $x$, \emph{even if the
adversary later learns the shared key that was used to derive $z$}
\cite{AumannDR:boundedstorage,DzieMaurer:boundedstorage}. Viewing our
communication result in the bounded-storage context, Alice's message is the
storage of the adversary, while Bob's input takes the role of the secret key
$K$. Our result shows that $z$ can fairly safely be used as a key if the adversary has
less than $\sqrt{n}$ classical storage, while it is completely insecure if the
adversary has $\sqrt{n}$ (or even only polylogarithmic) \emph{quantum} storage.%
\footnote{In a way our result can be viewed as a strong extractor, albeit a
rather bad one, where the random seed (the sequence of edges of Bob's input, described
below) takes about $n$ bits.}

Finally, let us point out that both results can be modified to give a
separation in the simultaneous message passing model between the models of
classical communication with shared entanglement and classical communication
with shared randomness. Earlier, such a separation was known only for a
relational problem~\cite{gkrw:identification}, not for a Boolean function.

\section{The problem and its quantum and classical upper bounds}

We assume basic knowledge of quantum computation~\cite{nielsen&chuang:qc} and
(quantum) communication complexity~\cite{kushilevitz&nisan:cc,wolf:qccsurvey}.
The partial Boolean function that will give the separation is the following
problem, parametrized by a value $\alpha\leq 1/4$. It's a modification of the
Boolean Hidden Matching Problem from~\cite{bjk:q1way}.

\medskip

\noindent
{\bf Alice:} $x\in\01^n$\\
{\bf Bob:} $\alpha n$ disjoint edges $e_1=(i_1,j_1),\ldots,e_{\alpha
n}=(i_{\alpha n},j_{\alpha n})$
from ${[n]\choose 2}$ and a string $w\in\01^{\alpha n}$\\
Define $z_\ell=x_{i_\ell}\oplus x_{j_\ell}$ and $z=z_1\ldots z_{\alpha n}$\\
{\bf Promise:} $w=z\oplus b^{\alpha n}$ for a bit $b$\\
{\bf Function value:} $b$

\medskip

There is an easy $O(\log(n)/\alpha)$-qubit protocol for this problem that gives the
correct output with probability 1/2 and claims ignorance otherwise, as follows. Given a
uniform superposition over all bits of $x$ (which takes $\log n$ qubits), Bob
can complete his $\alpha n$ edges to a perfect matching and measure with the
corresponding set of $n/2$ 2-dimensional projectors.
With probability $2\alpha$ he will get one of the edges $e_\ell=(i_\ell,j_\ell)$
of his input. The state will then collapse to $(-1)^{x_{i_\ell}}\ket{i_\ell}+(-1)^{x_{j_\ell}}\ket{j_\ell}$,
from which Bob can obtain the bit $z_\ell=x_{i_\ell}\oplus x_{j_\ell}$ with certainty.
XORing this bit with the corresponding bit $w_\ell$ in his string $w$ gives the function value $b$.
The protocol gives Bob $O(1/\alpha)$ copies of the $\log n$-qubit state, so he learns $b$
with good probability (and knows when he doesn't).

The following is an easy \emph{classical} upper bound.
Suppose Alice uniformly picks a subset of $d\approx\sqrt{n/\alpha}$ of the bits
of $x$ and sends those to Bob.  It is easy to see by the birthday paradox that
now with high probability Bob will have both endpoints of at least one of his
$\alpha n$ edges. This enables him to compute the function value $b$. To send a uniform subset of $d$ bits from $x$, Alice would need to send about $d\log n$ bits to Bob, since she needs to describe the $d$ indices as well as their bitvalues.
However, by Newman's theorem~\cite{newman:random}, Alice can actually restrict her random choice to picking one out of $O(n)$ possible $d$-bit subsets, instead of one out of all ${n\choose d}$ possible subsets. Hence $d+O(\log n)$ bits of communication suffice.

In the sections below we show that this classical upper bound is essentially optimal
for $\alpha\approx 1/\sqrt{\log n}$, which gives the exponential quantum-classical separation.

\section{Strategy of the proof}

We prove a lower bound on classical communication with shared randomness for
the problem of the previous section. By the Yao principle, it suffices to
prove a lower bound for \emph{deterministic} protocols under the uniform input
distribution on the $x$'s, the edges, and $b$ (note that this fixes Bob's
second input $w$). Suppose we have a classical deterministic one-way protocol
with $c$ bits of communication and error probability at most $1/10$ under this
distribution. This protocol partitions the set of $2^n$ $x$'s into $2^c$ sets
$A_1,\ldots,A_{2^c}$, one for each possible message. At least half of the $x$'s
must occur in sets of size at least $2^{n-c-1}$, since the smaller sets
together contain fewer than $2^c\cdot 2^{n-c-1}=2^{n-1}$ $x$'s. Hence there
must be at least one set $A$ that contains at least $2^{n-c-1}$ $x$'s and has
error at most $1/5$, otherwise the overall error would be larger than $1/10$.
Hereafter we will analyze this set $A$.

From Bob's point of view the following happens when he receives the message
corresponding to $A$: $\alpha n$ disjoint edges $(i_\ell,j_\ell)$,
$\ell\in[\alpha n]$, uniformly picked from ${[n]\choose 2}$ are given, and an
unknown $x$ is picked uniformly from $A$. As before, let
$z_\ell=x_{i_\ell}\oplus x_{j_\ell}$ and $z=z_1\ldots z_{\alpha n}$. Bob needs
to figure out whether his second input equals $z\oplus 0^{n/4}$ or $z\oplus
1^{n/4}$. We will use capital letters to denote the corresponding random
variables. Our goal here is to show that $Z$ is close to uniformly distributed
when the edges are known but $x$ is not. Suppose we can show that if the
communication $c$ is ``small'', then $Z$ is more or less uniform: the total
variation distance is $d(Z,U_{\alpha n})=\frac{1}{2}\sum_{z\in\01^{\alpha
n}}|\Pr[Z=z]-2^{-\alpha n}|\leq \delta$ for some small $\delta$ (this is the
bulk of the proof below).\footnote{\label{tvdfootnote}Note that we include a
factor of $1/2$ in our definition of total variation distance.  This means that
the distance lies in the interval $[0,1]$, and if distributions $P$ and $Q$
have distance $\delta$, the probability of any event cannot change by more than
$\delta$, i.e., $|\Pr_P[E]-\Pr_Q[E]|\leq\delta$ for any event $E$.} Then also
$d(Z\oplus 0^{\alpha n},U_{\alpha n})\leq\delta$ and $d(Z\oplus 1^{\alpha
n},U_{\alpha n})\leq\delta$, and hence by the triangle inequality
$$
d(Z\oplus 0^{\alpha n},Z\oplus 1^{\alpha n})\leq d(Z\oplus 0^{\alpha
n},U_{\alpha n})+d(Z\oplus 1^{\alpha n},U_{\alpha n})\leq 2\delta.
$$
But distinguishing between the two distributions $Z\oplus 0^{\alpha n}$ and
$Z\oplus 1^{\alpha n}$ is exactly what Bob needs to do to determine $b$.  It is
well known that distinguishing between two distributions with variation
distance $2\delta$ can be done with probability at most $1/2+\delta$.
Accordingly, if $c$ is ``small'' then the success probability will be close to
$1/2$. Conversely, since Bob's success probability on the set $A$ is at least
$4/5$, $c$ must have been large.

\section{How biased are the bits of $Z$?}

We will analyze the distribution of $Z$, which depends on the known edges
$e_1=(i_1,j_1),\ldots,e_{\alpha n}=(i_{\alpha n},j_{\alpha n})$ as well as the
unknown $x\in A$. Intuitively, if $c$ is small (i.e.~$A$ is large), for most
strings $z\in\01^n$ we should have $\Pr[Z=z]\approx 2^{-\alpha n}$ and hence
$d(Z,U_{\alpha n})$ is small. Proving this will be quite technical.

We view the edges as being picked one by one. Since $A$ is quite
large, for most $(i,j)$-pairs roughly equally many $x$'s should have $x_i\oplus x_j=1$
as have $x_i\oplus x_j=0$. Thus we expect the first bit $Z_1$ to be close to
uniformly distributed when $x$ is picked uniformly from $A$. Similarly, we
would like the later bits $Z_\ell$ to be more or less uniform when conditioned
on values $Z_1=z_1,\ldots,Z_{\ell-1}=z_{\ell-1}$ for the earlier edges. More
formally, once $(i_1,j_1),\ldots,(i_{\ell-1},j_{\ell-1})$ and
$z_1,\ldots,z_{\ell-1}$ have been fixed, we define the ``$\ell$th bias'' by
$$
\beta_\ell=\Pr_{x\in A}[Z_\ell=1\mid Z_1=z_1,\ldots,Z_{\ell-1}=z_{\ell-1}]-1/2.
$$
This is a random variable, depending on the choice of $(i_\ell,j_\ell)$.
It is positive if $Z_\ell$ is biased towards 1, and negative if $Z_\ell$ is biased towards 0.

Fixing the first $\ell-1$ edges and conditioning on their bitvalues
$Z_1=z_1,\ldots,Z_{\ell-1}=z_{\ell-1}$ will shrink the set of possible $x$'s.
Let $A_\ell$ be the subset of $A$ that is still consistent.
Initially we have $|A_1|=|A|\geq 2^{n-c-1}$. When we pick the next edge $(i_\ell,j_\ell)$ and
its value $z_\ell$, the new set $A_{\ell+1}$ will be smaller by a factor
$1/2+\beta_\ell$ if $z_\ell=1$ and by a factor $1/2-\beta_\ell$ if $z_\ell=0$.
Hence we expect the set to shrink by a factor of about two for each new edge
and bitvalue for that edge, i.e., $|A_\ell|\geq 2^{n-c-\ell}$.
We have
$$
|A_\ell| = |A|\cdot \Pr[Z_1=z_1,\ldots,Z_{\ell-1}=z_{\ell-1}]
         = |A|\cdot \prod_{i=1}^{\ell-1}\left(1/2-(-1)^{z_i}\beta_i\right),
$$
and in particular
$$
\Pr[Z=z]=\prod_{\ell=1}^{\alpha n}\Pr[Z_\ell=z_\ell\mid
Z_1=z_1,\ldots,Z_{\ell-1}=z_{\ell-1}]=\prod_{\ell=1}^{\alpha
n}\left(1/2-(-1)^{z_\ell}\beta_\ell\right)=\frac{|A_{\alpha n+1}|}{|A|}.
$$
Hence showing that $Z$ is close to uniformly distributed is equivalent to showing
that $|A_{\alpha n+1}|/|A|\approx 2^{-\alpha n}$ with high probability.

We use a result of Talagrand~\cite{talagrand:increasingsets} to relate the
expected squared bias $\beta_\ell^2$ to the size of the set $A_\ell$.

\begin{lemma}[\cite{talagrand:increasingsets}, Eq.~(2.9)]\label{lemtalagrand}
There is an absolute constant $K\geq 1$ such that for all $A \subseteq
\{0,1\}^n$
$$
\sum_{i,j\in[n],i \neq j} \beta_{ij}^2\leq K \left(\log \frac{K 2^n}{|A|}\right)^2,
$$
where $\beta_{ij}=\Exp_{x \in A} [x_i \oplus x_j-1/2]$.
\end{lemma}

This will allow us to establish a bound showing that $\beta_\ell$ is probably
quite small \emph{if} the set $A_\ell$ hasn't shrunk too fast. We allow some
more shrinking than we expect:  note the `$3 c$' instead of `$c$' in the
exponent below.

\begin{corollary}\label{lembeta}
There is an absolute constant $\gamma >0$ such that if $|A_\ell|\geq 2^{n-3c-\ell}$, then\\
(1) $\displaystyle \Exp[\beta_\ell^2]\leq \gamma ( c/n)^{2}$ and
(2) $\Pr [|\beta_\ell| \geq \eps]\leq \gamma(\frac{ c}{n \eps})^2$.
\end{corollary}

\begin{proof}
Note that fixing a bitvalue for the parity of an edge means that the two bits
in that edge behave as one bit. Accordingly, we can view the set $A_\ell$ as a
set of strings of length $m=n-(\ell-1)$ bits. We can upper bound the sum of
biases over all possible new edges (excluding ones touching earlier edges) by
the sum over all possible edges (including ones touching earlier edges):
$$
\sum_{i_\ell
j_\ell\not\in\{i_1,\ldots,i_{\ell-1},j_1,\ldots,j_{\ell-1}\}}\beta_{i_\ell,j_\ell}^2
\leq \sum_{i,j\in[m],i \neq j}\beta_{ij}^2\leq O(c^2),
$$
where the last inequality is by applying Lemma~\ref{lemtalagrand} to $A_\ell$.
Dividing by the number ${n-2(\ell-1)\choose 2}=\Theta(n^2)$ of possible new edges proves
part (1). Part (2) now follows from Chebyshev's inequality.
\end{proof}

There is a threat of circularity in our proof. On the one hand we need to assume that the sets $A_\ell$ are not too small in order to show that the biases $\beta_\ell$ are not too large (via Corollary~\ref{lembeta}).  But on the other hand we need to show that the biases are not too large in order to be able to conclude that $A_\ell$ is not too small.  To deal with this problem, below we give a proof in two ``passes''.  The first pass is quite coarse-grained and shows that (with high probability) the sets $A_\ell$ won't shrink by a factor of $2^{-2c}$ more than what we expect. This allows us to apply Corollary~\ref{lembeta} to each of the $\alpha n$ biases during the second pass.  In this second, more fine-grained pass we actually show that $d(Z,U_{\alpha n})$ is small.

\section{First pass: The sets $A_\ell$ probably don't shrink too much}

We can only use Corollary~\ref{lembeta} if the condition $|A_\ell|\geq 2^{n-3c-\ell}$ is satisfied.
We now show that with high probability this is indeed
the case for each $\ell$ simultaneously. The proof uses the following concentration result
from~\cite{McDiarmid:concentration}.

\begin{lemma}[\cite{McDiarmid:concentration}, special case of Thm.~3.7]\label{McDiarmid}
Let $S_1,\ldots,S_k$ be bounded random variables satisfying $\Exp[S_j|S_{1}=s_{1},\ldots,S_{j-1}=s_{j-1}]=0$
for all $1 \leq j\leq k$ and all values $s_{1},\ldots,s_k$.
Then for all $t,v \geq 0$
$$
\Pr\left[\sum_{j=1}^k S_j \geq t\right]\leq e^{-t^2/2v}+\Pr\left[\sum_{j=1}^k
S_j^2\geq v\right].
$$
\end{lemma}

\begin{lemma}\label{lemAs}
There is a constant $\delta_0 >0$ such that for all $0<\delta \leq \delta_0$,
if $\alpha=\delta^2/(4 \sqrt{\ln n})$ and $c =\delta \sqrt{n}(\ln n)^{1/4}$
then with probability $99/100$ over all choices of $e_1,\ldots,e_{\alpha n}$
and $z=z_1,\ldots,z_{\alpha n}$ the following holds: for each $\ell\in[\alpha
n]$ we have $|A_\ell|\geq 2^{n-3c-\ell}$ and $|\beta_\ell| \leq 1/4$.
\end{lemma}

\begin{proof}
Note that
$$
|A_\ell|  = |A|\cdot \Pr[Z_1=z_1,\ldots,Z_{\ell-1}=z_{\ell-1}]
       \  = |A|\cdot \prod_{i=1}^{\ell-1}\left(1/2-(-1)^{z_i}\beta_i\right)
       \geq  2^{n-c-\ell}\prod_{i=1}^{\ell-1}\left(1-(-1)^{z_i}2\beta_i\right).
$$
Define $S_i=-(-1)^{z_i}2 \beta_i$. For the lower bound on $A_\ell$ it thus suffices to lower bound
$\prod_{i=1}^{\ell-1}(1+S_i)$ by $2^{-2c}$. Taking logarithms, we need to show
for each $\ell$
\begin{equation}\label{eq:toshow}
\sum_{i=1}^{\ell-1}\log(1+S_i)\geq -2c.
\end{equation}
Let us divide the $\alpha n$ $\ell$s into blocks of size $c$, i.e., for $1 \leq
k \leq \alpha n/c$ define the $k$th block $B_k=\{(k-1) c+1,\ldots,{k c }\}$ (we ignore rounding for simplicity).
Let $E_k$ be the following event:
\begin{quote}
(a) $|\beta_i| \leq 1/4$ for each $i \in B_k$  and\\
(b) $\sum_{i \in B_k} \log (1+S_i) \geq - c^2/\alpha n$.
\end{quote}
We will show below in Claim~\ref{claim1} that for all $k$, $\Pr[\neg E_k\mid E_1,\ldots,E_{k-1}] \leq c/100
\alpha n$. This implies
$$
\Pr[\neg(E_1,\ldots,E_{\alpha n/ c})] \leq \sum_{k=1}^{\alpha n/ c} \Pr[\neg
E_k\mid E_1,\ldots,E_{k-1}]\leq \frac{\alpha n}{ c} \cdot  \frac{c}{100 \alpha
n}=\frac{1}{100}.
$$
If $E_1,\ldots,E_{\alpha n/ c}$ all hold, then from (b) for all $k$ we have
$\sum_{i=1}^{k \cdot  c} \log(1+S_i)\geq -k \cdot c^2/\alpha n\geq -c$ and in
particular Eq.~(\ref{eq:toshow}) holds whenever $\ell-1$ is a multiple of $c$.
For the other $\ell$ pick $k$ such that $\ell-1 \in B_{k+1}$ and note that
thanks to (a) we have $\log(1+S_i)\geq \log(1-2(1/4))= -1$ and
hence
$$
\sum_{i=1}^{\ell-1}\log(1+S_i)=\sum_{i=1}^{kc}\log(1+S_i)+\sum_{i=kc+1}^{\ell-1}\log(1+S_i)
\geq -c+\sum_{i=kc+1}^{\ell-1}-1 \geq -2c.
$$

\begin{claim}\label{claim1}
 For all $1 \leq k
\leq \alpha n/c$, we have $\Pr[\neg E_k\mid E_1,\ldots,E_{k-1}] \leq  c/100 \alpha
n$.
\end{claim}

\begin{proof}
Let $\ell_1=(k-1)c$ be the last index in $B_{k-1}$ and condition on $E_1,\ldots,
E_{k-1}$. This means that $|A_{\ell_1+1}|\geq 2^{n-2c-\ell_1-1}$. Let $F_i$
denote the event that (a) holds for $i \in B_k$, i.e., $|\beta_i|\leq 1/4$. We
want to show $\Pr[\neg F_{i}\mid F_1,\ldots,F_{i-1}]\leq 1/500 \alpha n$
for $i \in B_k$. This will imply that (a) fails to hold only with probability at
most $c/500 \alpha n$. If $|\beta_{\ell_1+1}|,\ldots,|\beta_{\ell_1+i-1}|\leq
1/4$ then as before $\sum_{j=\ell_1+1}^{i-1}\log(1+S_j) \geq -c$ and hence
$|A_{\ell_1+i}|\geq |A_{\ell_1+1}|\cdot 2^{-i-c}\geq 2^{n-3c-(\ell_1+i)}$. We can now
apply Corollary~\ref{lembeta}(part 2) to show
$$
\Pr[|\beta_{\ell_1+i}|>1/4]\leq \gamma
\left(\frac{4 c}{n}\right)^2=16 \gamma \delta^2 \frac{\sqrt{\ln n}}{n} \leq
\frac{1}{500 \alpha n},
$$
where we use $\alpha=\delta^2/(4 \sqrt{\ln n})$ and
$c = \delta \sqrt{n}(\ln n)^{1/4} $ and choose $\delta_0$ small enough.

Now, assuming (a) holds for each $i \in B_k$ and hence the conditions of
Corollary~\ref{lembeta} hold for each $i \in B_k$, we will show that (b) holds
for $B_k$ with probability at least $1-4 c/500 \alpha n$, which will imply the
claim.

Note that $\log(1+S_i) \geq S_i-2S_i^2$ if $|\beta_i| \leq 1/4$ and hence
\begin{equation}\label{eqnsumsi}
\sum_{i \in B_k}\log(1+S_i)\geq \sum_{i \in B_k} S_i - 2\sum_{i \in B_k}S^2_i=\sum_{i \in B_k} S_i - 8\sum_{i \in B_k}\beta^2_i.
\end{equation}
We first bound the second term of the righthand side.
Corollary~\ref{lembeta} implies
$$
\Exp\left[\sum_{i \in B_k} \beta^2_i\right] \leq c \cdot \gamma(c/n)^2=\gamma c^3/n^2.
$$
Let $v=c^2/2\alpha n$; this is half of what (b) allows us to lose.  By Markov's inequality
$$
\Pr\left[8\sum_{i \in B_k} \beta_i^2 >  v\right]
\leq  16 \alpha \gamma c/n \leq c/500 \alpha n
$$
for sufficiently large $n$.

Now for the first term in the righthand side of Eq.~(\ref{eqnsumsi}).
Conditioning on event (a) changes the (a priory uniform) distribution on the
$z_i$ for the $i \in B_k$ by at most $c/500 \alpha n$ in total variation
distance. This means that if we bound $\Pr[\sum_{i \in B_k} S_i \leq -v]$ under
the assumption that the $z_i$ are uniform, the true probability will change by
at most $c/500 \alpha n$ (see footnote~\ref{tvdfootnote}). \emph{If} the $z_i$
are uniform then the condition of Lemma~\ref{McDiarmid} holds for each $S_i$:
the conditional expectations are all 0, because the sign of $S_i$ is $+$ or $-$
with equal probability. Recall that $S_i^2=4\beta_i^2$ and $v=c^2/2\alpha n$.
Hence by Lemma~\ref{McDiarmid} (with $t=v$), if the $z_i$ are uniform
$$
\Pr\left[\sum_{i \in B_k} S_i < -v\right]\leq e^{-v/2}+\Pr\left[4 \sum_{i \in
B_k} \beta^2_i \geq v\right]\leq 1/n + c/500\alpha n \leq 2c/500\alpha n.
$$
Putting everything together we upper bound the probability that (b) fails for the $k$th block:
$$
\Pr\left[\sum_{i \in B_k} \log(1+S_i)< -c^2/\alpha n=-2v\right] \leq
\Pr\left[\sum_{i \in B_k}S_i < -v\right]+
\Pr\left[8\sum_{i \in B_k} \beta_i^2 > v \right]\leq 4c/500 \alpha n.
$$
This concludes the proof of Claim~\ref{claim1}.
\end{proof}

\noindent
This concludes the proof of Lemma~\ref{lemAs}.
\end{proof}

\section{Second pass: $Z$ is close to uniform}

We now prove the main result, which implies the $\tilde \Omega(n^{1/2})$ lower
bound on classical one-way communication.

\begin{theorem}
There is a constant $\delta>0$ such that if $c= \delta  \sqrt{n}(\ln n)^{1/4}$
and $\alpha=\frac{\delta^2}{4 \sqrt{\ln n}}$, then $d(Z,U_{\alpha n})\leq
1/10$.
\end{theorem}

Of course this theorem also holds if the communication of the classical protocol
is $c<\delta n^{1/2}(\ln n)^{1/4}$, since we
can always add dummy bits to a shorter message to make its length
equal to exactly that value.

\medskip

\begin{proof}
We rewrite the total variation distance:
\begin{eqnarray*}
 d(Z,U_{\alpha n}) & = & \frac{1}{2}\sum_{z \in \{0,1\}^{\alpha n}}\left| \Pr[Z=z] - 2^{-\alpha n}\right|\\
& = & \frac{1}{2}2^{-\alpha n}\sum_{z \in \{0,1\}^{\alpha n}}\left|\prod_{\ell=1}^{\alpha n}(1-(-1)^{z_\ell}2\beta_\ell)-1\right|\\
& = & \frac{1}{2}\Exp_z\left[\left|\prod_{\ell=1}^{\alpha n}(1-(-1)^{z_\ell}2\beta_\ell)-1\right|\right],
\end{eqnarray*}
where $\Exp_z$ denotes the expectation over uniform $z$. By Lemma~\ref{lemAs},
with probability $99/100$, for each $\ell\in[\alpha n]$ we have $|A_\ell|\geq
2^{n-3c-\ell}$ and $|\beta_\ell|\leq 1/4$. Let us call this event $E$. Then
$$
d(Z,U_{\alpha n}) \leq \Pr[E]\cdot d(Z|_E,U_{\alpha n})+\Pr[\neg E]\cdot d(Z|_{\neg
E},U_{\alpha n})\leq \frac{99}{100} \cdot d(Z|_E,U_{\alpha n})+\frac{1}{100}.
$$
Note that conditioning on $E$ will change the (a priori uniform) distribution
on the $z$. However, the total variation distance between the conditioned
distribution $z|_E$ of the $z$ and the uniform distribution is at most $1/100$,
since the event $E$ on which it is conditioned has probability at least $99/100$.
Hence
\begin{eqnarray*}
d(Z|_E,U_{\alpha n}) & = & \Exp_{z|_E}\left[\left|\prod_{\ell=1}^{\alpha n}(1-(-1)^{z_\ell}2 \beta_\ell)-1\right|\right]\\
 & \leq & \Exp_{z}\left[\left|\prod_{\ell=1}^{\alpha n}(1-(-1)^{z_\ell}2 \beta_\ell)-1\right|\right]+1/100\\
 & = & \Exp_z\left[\left|2^{\sum_{\ell=1}^{\alpha n}\log(1+S_\ell)}-1\right|\right]+1/100,
\end{eqnarray*}
where $S_\ell=-(-1)^{z_\ell}2 \beta_\ell$ as in the previous section.
We thus need to show that $\sum_{\ell=1}^{\alpha n}\log(1+S_\ell)$ is usually very close to 0.
When conditioned on event $E$ we have $|\beta_\ell|\leq 1/4$ and hence
$$
\frac{S_\ell}{\ln 2}\geq \log(1+S_\ell) \geq S_\ell-2S_\ell^2.
$$
It thus suffices to show that with high probability, both
$|\sum_{\ell=1}^{\alpha n}S_\ell |$ and $\sum_{\ell=1}^{\alpha n}S^2_\ell$ are
small. This can be done in the same way as in the proof of Claim~\ref{claim1},
using this time that $\Exp[\sum_{\ell=1}^{\alpha n} \beta_\ell] \leq \gamma
\alpha c^2/n=o(1)$.
%
\end{proof}

\section{Acknowledgments}

We thank Oded Regev and Guy Kindler for referring us to Talagrand's result, and
Jaikumar Radhakrishnan for a reference to McDiarmid's martingale bound. Many
thanks to Iordanis Kerenidis and Ran Raz for discussions and for delaying the
publication of their proof until ours was ready too. Thanks to Renato Renner
and Christian Schaffner for discussions about the bounded-storage model.

\bibliographystyle{alpha}

\end{document}